\documentclass[aps,prl,twocolumn,tightenlines,amsmath,amssymb,superscriptaddress]{revtex4-1}

\usepackage{graphicx}
\usepackage{amsmath}
\usepackage{amssymb}
\usepackage{epstopdf}
\usepackage{color}
\usepackage{lipsum}
\usepackage{ulem}
\usepackage[dvipsnames]{xcolor}
\usepackage{array}
\DeclareGraphicsExtensions{.eps}
\usepackage{hyphenat}

\newcommand{\Tr}{\text{Tr}} 

\newcommand{\bra}[1] {\langle #1|}
\newcommand{\ket}[1] {|#1 \rangle}

\newcommand{\ketbra}[2] {{|#1 \rangle\!\langle #2|}}

\newcommand{\sigp}[1] {R_#1}
\newcommand{\sigpp} {R}
\newcommand{\sigm}[1] {L_#1}
\newcommand{\sigmm} {L}
 
\newcommand*{\rom}[1]{\expandafter\@slowromancap\romannumeral #1@}

\newcommand{\fixtable}[2]{\parbox[m]{#1}{\vspace{0.2em} #2 \vspace{0.2em}}}

\begin{document}

\title{High-rate entanglement between a semiconductor spin and indistinguishable photons }

\author{N. Coste}
\author{D. Fioretto}
\affiliation{Universit\'e Paris-Saclay, CNRS, Centre de Nanosciences et de Nanotechnologies, 91120, Palaiseau, France}
\author{N. Belabas}
\affiliation{Universit\'e Paris-Saclay, CNRS, Centre de Nanosciences et de Nanotechnologies, 91120, Palaiseau, France}
\author{S. C. Wein}
\affiliation{Quandela SAS, 10 Boulevard Thomas Gobert, 91120, Palaiseau, France}
\affiliation{Universit\'{e} Grenoble Alpes, CNRS, Grenoble INP, Institut N\'eel, 38000 Grenoble, France}
\author{P. Hilaire}
\affiliation{Huygens-Kamerlingh Onnes Laboratory, Leiden University, P.O. Box 9504, 2300 RA Leiden, The Netherlands}
\affiliation{Quandela SAS, 10 Boulevard Thomas Gobert, 91120, Palaiseau, France}

\author{{R. Frantzeskakis}}
\affiliation{Department of Physics, University of Crete, Heraklion, 71003, Greece}
\author{M. Gundin}
\affiliation{Universit\'e Paris-Saclay, CNRS, Centre de Nanosciences et de Nanotechnologies, 91120, Palaiseau, France}
\author{B. Goes}
\affiliation{Universit\'{e} Grenoble Alpes, CNRS, Grenoble INP, Institut N\'eel, 38000 Grenoble, France}

\author{N. Somaschi}
\affiliation{Quandela SAS, 10 Boulevard Thomas Gobert, 91120, Palaiseau, France}
\author{M. Morassi}
\affiliation{Universit\'e Paris-Saclay, CNRS, Centre de Nanosciences et de Nanotechnologies, 91120, Palaiseau, France}
\author{A. Lema\^itre}
\author{I. Sagnes}
\author{A. Harouri}
\affiliation{Universit\'e Paris-Saclay, CNRS, Centre de Nanosciences et de Nanotechnologies, 91120, Palaiseau, France}
\author{{S. E. Economou}}
\affiliation{Department of Physics, Virginia Tech, Blacksburg, Virginia, 24061, USA}

\author{A. Auffeves}
\affiliation{Universit\'{e} Grenoble Alpes, CNRS, Grenoble INP, Institut N\'eel, 38000 Grenoble, France}
\author{O. Krebs}
\affiliation{Universit\'e Paris-Saclay, CNRS, Centre de Nanosciences et de Nanotechnologies, 91120, Palaiseau, France}

\author{L. Lanco}
\affiliation{Universit\'e Paris-Saclay, CNRS, Centre de Nanosciences et de Nanotechnologies, 91120, Palaiseau, France}
\affiliation{Universit\'e Paris Cit\'e, CNRS, Centre de Nanosciences et de Nanotechnologies, 91120, Palaiseau, France}
\author{P. Senellart}
\affiliation{Universit\'e Paris-Saclay, CNRS, Centre de Nanosciences et de Nanotechnologies, 91120, Palaiseau, France}


	\begin{abstract}
	 	Photonic graph states, quantum light states where multiple photons  are mutually entangled, are key resources for optical quantum technologies. They are {notably} at the core of error-corrected measurement-based optical quantum computing~\cite{raussendorf_measurement-based_2003,Raussendorf2005,Raussendorf2007,fault-tolerant-2020,Rudolph2017} and all-optical quantum networks~\cite{Azuma2015}. {In the discrete variable framework,} these applications require {high efficiency generation of cluster-states whose nodes are indistinguishable photons}. {Such} photonic cluster states can be generated with heralded single photon sources and probabilistic quantum gates, yet with {challenging} efficiency and scalability~\cite{Zhong2018}. Spin-photon entanglement has been proposed to deterministically generate linear cluster states {\cite{Schon2005,lindner_proposal_2009}}. First demonstrations have been obtained with semiconductor spins~\cite{schwartz_deterministic_2016} achieving high photon indistinguishablity~\cite{CoganClusterHole2021},  and most recently with atomic systems~\cite{JWP2021cluster,RempeCluster2022} at  high collection efficiency and  record length~\cite{RempeCluster2022}. Here we report on the efficient generation of three partite cluster states made of {one} semiconductor spin and two indistinguishable photons. We {harness a} semiconductor quantum dot inserted in an optical cavity for efficient photon collection and {electrically controlled} for high indistinguishability. We demonstrate two and three particle entanglement with fidelities of { 80 \%} and {63 \%} respectively,  with photon indistinguishability of 88\%. The spin-photon and spin-photon-photon entanglement rates exceed by  three and  two orders of magnitude respectively the previous state of the art~\cite{RempeCluster2022}. {Our system  and experimental scheme,} a monolithic solid-state device controlled with a resource efficient simple experimental configuration, are very promising for future {scalable} applications. 
	 	\end{abstract}

\maketitle

Measurement-based quantum computing and quantum networks have been proposed to overcome the difficulty of implementing quantum logical gates between single photons~\cite{raussendorf_measurement-based_2003,Azuma2015}. Both rely on the engineering of dedicated multi-photon entangled 
{states} such as GHZ~\cite{Greenberger1989} or cluster states~\cite{Briegel2001}. In this approach, the entanglement structure between the photons allows implementing multi-qubit 
logical gates  performing only single qubit gates and measurements, {which are both} straightforward to implement with {optical platforms}. Moreover, the redundancy of the entanglement structure {is at the core of } error correction and loss mitigation~\cite{Raussendorf2005,Raussendorf2007,fault-tolerant-2020,Rudolph2017,Zhang2022}. 

Photonic cluster states can be 
{created using} single photon sources and linear{-optical}  gates~\cite{Zhong2018}. 
Given the intrinsically low efficiency of {parametric} photon sources~\cite{Senellart2017} and the probabilistic nature of the gates~\cite{Knill2001}, 
{scaling up the number of nodes i.e. photons in} such a scheme requires {extensive} integration to make the source efficient and many ancilla photons to  herald the gates~\cite{Knill2002}. {More tractable} scalability was demonstrated with a bright single photon source based on semiconductor quantum dots (QDs) {and} a probabilistic gate in a resource efficient time loop entangling apparatus~\cite{istrati_sequential_2020}. Yet, the most efficient schemes  so far have been {theoretically} proposed in references~\cite{Schon2005,lindner_proposal_2009} where linear cluster states are deterministically generated exploiting the optical selection rules  {that} connect a spin state to the polarization of  emitted single photons. Successive {pulsed and timed excitations} of the system lead to the generation of a {train} of photons {that are all} entangled with the same spin. This scheme only allows for the generation of one-dimensional cluster states {whereas} higher dimensionality is required for fault-tolerant quantum computing protocols. Linear optical gates {can however yield} higher dimension states provided the single photons in the linear cluster states are indistinguishable~\cite{Knill2001,Browne2005}. 

\noindent {After the demonstration of spin-photon entanglement~\cite{SpinPhoton1,SpinPhoton2,SpinPhoton3}, t}he generation of three partite cluster states was first reported for a  QD using a dark exciton spin~\cite{schwartz_deterministic_2016}. More recently, a hole spin in a QD was exploited allowing for high indistinguishability of the photons~\cite{CoganClusterHole2021}. In both cases, the source efficiency was limited to less than 1\% due to the absence of  photonic structure needed for efficient collection. Impressive {progresses have} recently {been} reported with macroscopic atomic systems in a cavity, leading to 6 photon entanglement with a Rydberg super-atom~\cite{JWP2021cluster} and a record efficiency for up to 14 photon-entanglement with a single atom in a cavity~\cite{RempeCluster2022}. Yet, to the best of our knowledge, an integrated bright solid-state source of indistinguishable photons in a linear cluster state is still lacking. {This is what we address in the present work based on} a semiconductor QD spin deterministically integrated in a monolithic cavity structure.

\begin{figure}
	\includegraphics[width=0.85\linewidth]{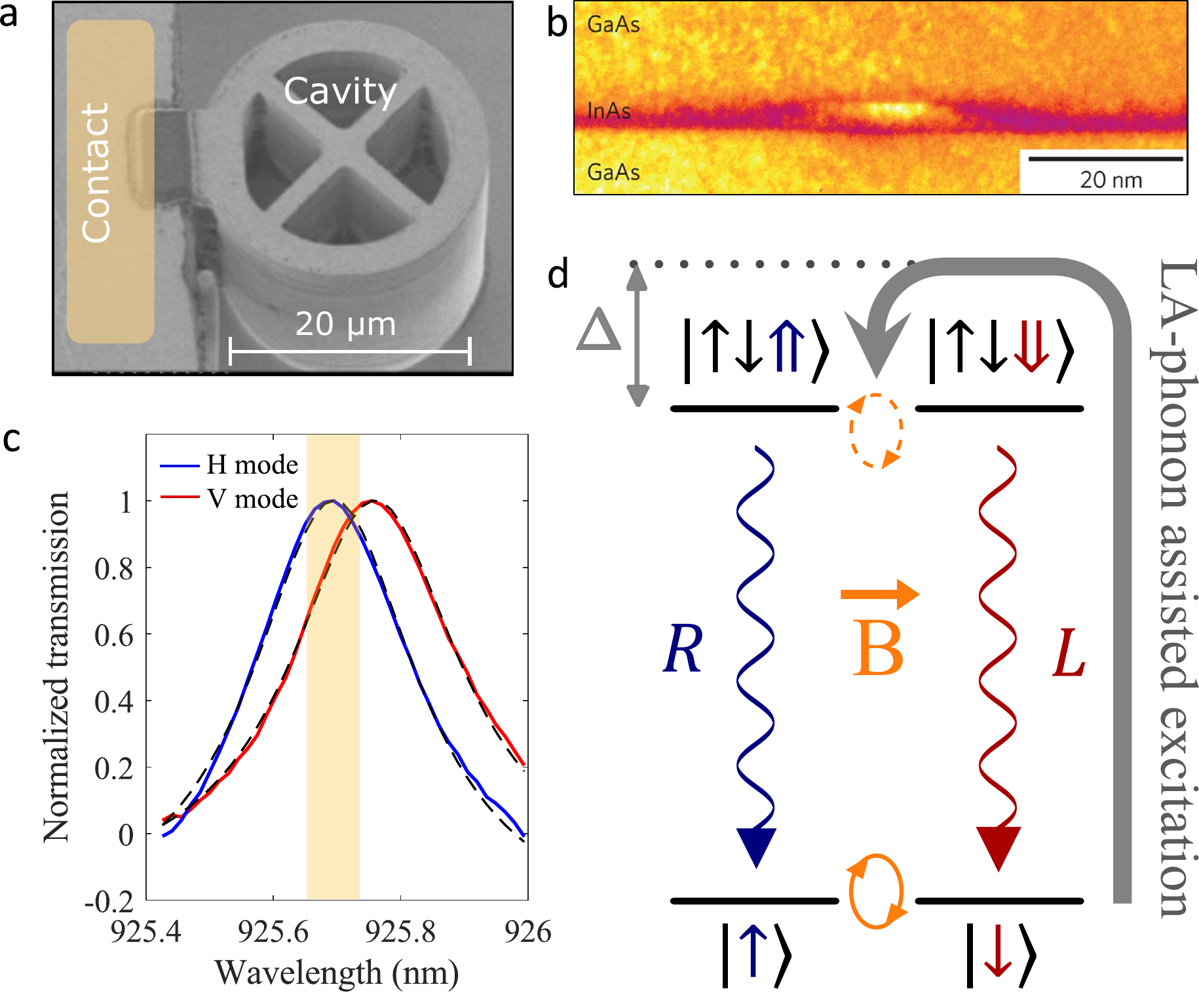}
	\caption{ { \textbf{A monolithic solid-state spin-photon interface.} a. Scanning electron microscopy image of 
	a 20~$\mu$m connected pillar optical cavity~\cite{Nowak2014} 
	In its center, a single negatively charged QD  b. Transmission electron microscopy image of a single InGaAs QD in GaAs barriers. 
	c. { Normalised transmission spectra of the two linearly polarized fundamental cavity modes obtained from reflectivity measurements. }
	The orange shaded area corresponds to the energy of the trion transition that can be adjusted with {an} applied bias. d. Energy levels and optical selection rules of the negatively charged QD in the presence of a small {($<100\ $mT} transverse magnetic field). The excitation laser is $\Delta=$0.8~nm blue detuned from the transition to implement {selection rule-preserving} acoustic-phonon assisted excitation.
	} \label{Fig_1}}
\end{figure}

Our semiconductor device is shown in the scanning-electron microscope image of Fig.\ref{Fig_1}.a. It consists of a wheel-shaped microcavity which confines light in an optical volume of the order of few $\lambda^3$ with a single InGaAs QD positioned in its center with nanometer scale accuracy owing to the in-situ lithography technique~\cite{Dousse2008}. The QD is inserted in a n-i-p diode structure and the cavity is connected to an electrical contact that allows applying an electrical bias to tune the QD transitions and reduce charge noise~\cite{somaschi_near-optimal_2016}. Fig.\ref{Fig_1}.b. presents the energy levels and optical selection rules of the QD charged with a single electron. In the absence of applied magnetic field, the two ground states correspond to the spin $|\uparrow \rangle$ and  $|\downarrow \rangle$ states. The excited states correspond to trion transitions where two electrons and one hole are either in  the $|\uparrow \downarrow \Uparrow \rangle$ or in the $|\uparrow \downarrow \Downarrow \rangle$ state. The symmetry of the system leads to two optical transitions  corresponding to circularly polarized right (R) or left (L) emitted photon.

To generate cluster states, a {weak} ($B\approx 40$ mT) in-plane magnetic field (Voigt configuration) is applied, leading to a precession of the spin with a Larmor period $T_e=2\pi / \omega_e=2 \pi \hbar[g_\mathrm{e} \mu_\mathrm{B} B]^{-1}=3.25$~ns, much slower than the typical time of the optical transition ($\approx200\ $ps). The cluster-state generation scheme exploits the spin-photon entanglement arising from the selection rules connecting the ground spin states with the photon polarization {(Fig.\ref{Fig_1}.d)}. To efficiently collect the single photons emitted by the QD, we exploit the Purcell effect,  where the cavity mode emission rate is accelerated compared to the emission in any other direction. An important requirement for  cluster state generation is to implement an unpolarized Purcell effect {although} nanofabrication processes and strain in the material usually lead to  bi-refringence. Fig.\ref{Fig_1}.c. presents the measured  cavity modes obtained through reflectivity measurements and the vertical shaded area highlights the energy range of the trion transition controlled via the external bias. {Adjusting} the {spectral} detuning of the transition with respect to these two modes  {enable} near-unpolarized photon extraction. A residual linear polarisation of $2-4\%$ is observed for the measurements presented below. To operate the entanglement protocol in the polarization bases, otherwise unaccessible via the standard resonant excitation regime, we use a longitudinal-acoustic (LA) phonon-assisted excitation scheme, where a 20 ps pulse excitation laser is blue-detuned  from the optical transition by approximately $\Delta\approx 0.8\ $nm (Fig.\ref{Fig_1}.d.). This has recently been shown to allow high occupation probability of the trion state and high photon indistinguishability~\cite{thomas_bright_2021} while preserving the same spin-selective optical selection rules as for resonant excitation~{\cite{Coste-precession-2022}. 

Fig.\ref{Fig_2}.a. presents the experimental schemes used to generate entanglement between the electron spin and the emitted photons. For clarity, we first explain  the protocol in an ideal and simplified case where both the emission lifetime  and spin decoherence are negligible. Three linearly polarized laser pulses (labelled 1, 2, and 3) are used to excite the QD with respective delays $t_{12}=t_2-t_1$ and $t_{23}=t_3-t_2$. The scheme  is repeated every $12\ $ns {at} the repetition rate of the laser (81~MHz). Considering a coherence time of the electron spin of a few nanoseconds similar to the precession period, we assume that it is in a perfectly mixed with equal probabilities for spin up and spin down state at $t_1$. 
The first pulse is then used to initialize the spin state. Detecting a first $\sigpp$ polarized photon (photon  $\#$1) {heralds a spin at $t_1$}  in the $|\uparrow \rangle$ state. Between pulses 1 and 2, the spin precesses around the in-plane magnetic field and $t_{12}$ is set so that  the spin  is in state  $|+ \rangle$ at $t_2$ where $|\pm \rangle=\frac{1}{\sqrt{2}}(|\uparrow \rangle\pm|\downarrow \rangle)$.  This corresponds to  $t_{12} = 810\pm 40$ps. After the emission process related to pulse 2, the system is ideally in the entangled state { $|\psi_2\rangle=\frac{1}{\sqrt{2}}(|\uparrow ,\sigp{2}\rangle+|\downarrow ,\sigm{2}\rangle)$
 { or equivalently
{$\ket{\psi_2}= \frac{1}{\sqrt{2}}(\ket{+,H_2} - i \ket{-, V_2})$}
 with V and H linear polarization states  defined as
 {$\ket{V} = \frac{1}{i\sqrt{2}}(\ket{\sigmm} - \ket{\sigpp})$}
 and
 {$\ket{H} = \frac{1}{\sqrt{2}}(\ket{\sigmm} + \ket{\sigpp})$}.
 During $t_{23}=t_{12}$, the state $|+\rangle$ ( resp. $|-\rangle$ ) evolves to
 {$|\downarrow\rangle$}
 {(resp. $|\uparrow\rangle)$}}
 so that after the  emission induced by the third excitation pulse, the system (spin, photon  $\#$2, photon $\#$3) is ideally in :
\begin{equation}
{\ket{\psi_3} = \frac{1}{\sqrt{2}}(-i |\uparrow ,V_2,\sigp{3}\rangle+|\downarrow ,H_2,\sigm{3}\rangle)
}\label{psi3}\end{equation} 
{\noindent a state locally equivalent to a GHZ or a three-qubit linear cluster}. Measuring the polarization of photon $\#$3 thus allows a direct measurement of the spin state at $t_3$. 
 
\begin{figure}[t]
	\includegraphics[width=\linewidth]{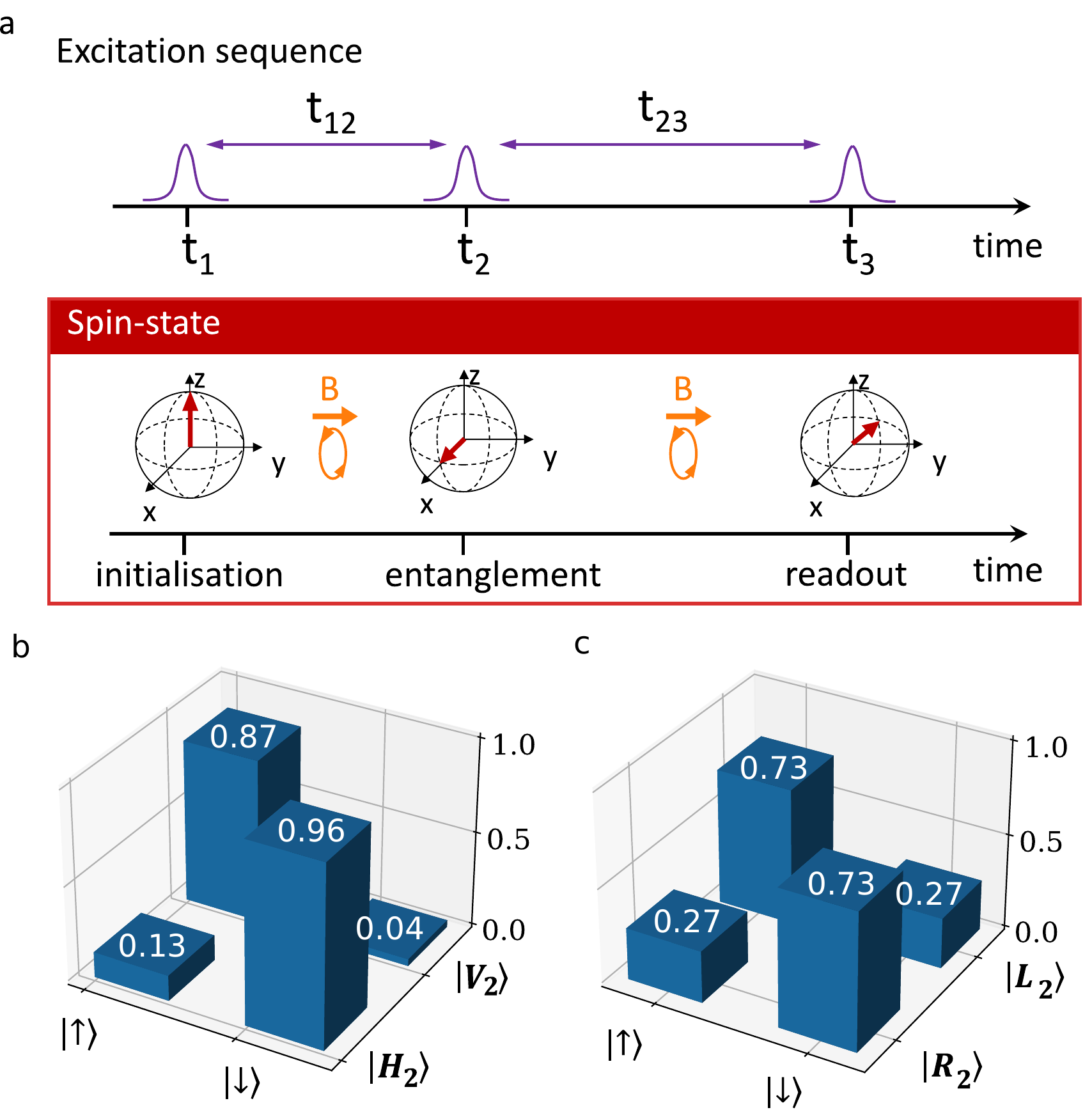}
	\caption{ { \textbf{Experimental scheme  and spin-photon entanglement}. a. Sketch of the {exciting} pulse sequence 
	and corresponding spin state. 
	{The first pulse is used to initialize the spin   in the $|\uparrow \rangle$ state. A second pulse is sent after a set $t_{12}$, such that the spin has precessed to a balanced superposition of $|\uparrow\rangle$ and $|\downarrow\rangle$. According to the selection rules of Fig.\ref{Fig_1}.d. the emitted photon $\#$2 is then entangled with the spin state. A third pulse is sent after a variable delay $t_{23}$ that leads to  spin-photon-photon entanglement and also allows for spin readout at $t_3$}. {Note that the sketch represents the spin state after initialisation  and precession in the absence of the entanglement generated by the two last pulses}.
	b and c. Truth tables representing the probability of the second photon polarization conditioned on the  spin state measured at time $t_3$, respectively after a delay $t_{23} =t_{12}= 810\pm 40$~ps 
 	 and $t_{23} =2t_{12}$.
	} \label{Fig_2}}
\end{figure}

}

{We first demonstrate spin-photon $\#$2 entanglement by measuring the spin state after the third excitation {by measuring photon $\#$3 in the R/L basis}. This is done for two delays $t_{23}$ to change the spin measurement basis.} Fig.\ref{Fig_2}.b. {gives the measured conditional probabilities of} photon $\#$2 in the $H_{2}$ or $V_{2}$ state and the spin in the $|\uparrow\rangle$ and  $|\downarrow\rangle$ states {at $t_3$} for $t_{23}=t_{12}$. This is obtained through three photon correlation measurements (initialisation, entanglement, spin read-out). We find 
{$P(V_{2}|\uparrow)=0.87\pm 0.02$, $P(H_{2}|\uparrow)=0.13{\pm 0.02}$, $P(V_{2}|\downarrow)=0.04{\pm 0.02}$ and $P(H_{2}|\downarrow)=0.96{\pm 0.02}$}. 


\noindent We then repeat the same experiment 
setting $t_{23}=2t_{12}$. {There, the system has ideally evolved to} the state at $t_3$: $${\frac{1}{\sqrt{2}}(|\downarrow , \sigp{2},\sigm{3}\rangle-|\uparrow ,\sigm{2},\sigp{3}\rangle)}.$$ \noindent Fig.\ref{Fig_2}.c. present the corresponding conditional probabilities
{$P(\sigp{2}|\uparrow)=0.27{\pm 0.02}$, $P(\sigm{2}|\uparrow)=0.73{\pm 0.02}$, $P(\sigp{2}|\downarrow)=0.73{\pm 0.02}$ and $P(\sigm{2}|\downarrow)=0.27{\pm 0.02}$.}
Using these two measurements, following~\cite{Blinov2004}, we find a lower bound for the fidelity $F$ to the spin-photon entangled state $|\psi_2\rangle$ of $F> {F_{s,p} =}65{\pm 1}\%$. The actual fidelity at $t_2$ is actually much larger since the spin readout process at $t_3$ is degraded  by the spin decoherence during the time interval $t_{23}$.


    From these measurements, that rely on the three partite entanglement of the spin with photons $\#$2 and $\#$3, we can derive a {fidelity lower bound to the three-partite spin-photon-photon entangled state (at $t_{23} = t_{12}$)}
    under two  assumptions that are experimentally fulfilled here (i) the spin is in perfect mixed state at the beginning of the sequence (ii) two successively emitted photons after spin initialisation have exactly the same polarization. 
    We find (see supplementary) a lower bound {$F_{s,2p}$} to  $\ket{\psi_3}$ at $t_{23} = t_{12}$: $$F_{s,2p} = F_{s,p} (1 - s_{X})/2 $$} where $F_{s,p}$ is the spin-photon fidelity lower bound and $s_{X}$ is the result of  a measurement on photon \#2 along the H/V axis {in the three-photon coincidence experiment at $t_{23} = t_{12}$}. {We find $F_{s,2p} = 59\%  {\pm 2} \%$, beyond the 50\% threshold for entanglement. }To the best of our knowledge, the present experimental measurements demonstrate for the first time genuine spin-photon-photon entanglement for a solid-state spin in a cavity. 
    
    
    \begin{figure}[t]
	\includegraphics[width=1\linewidth]{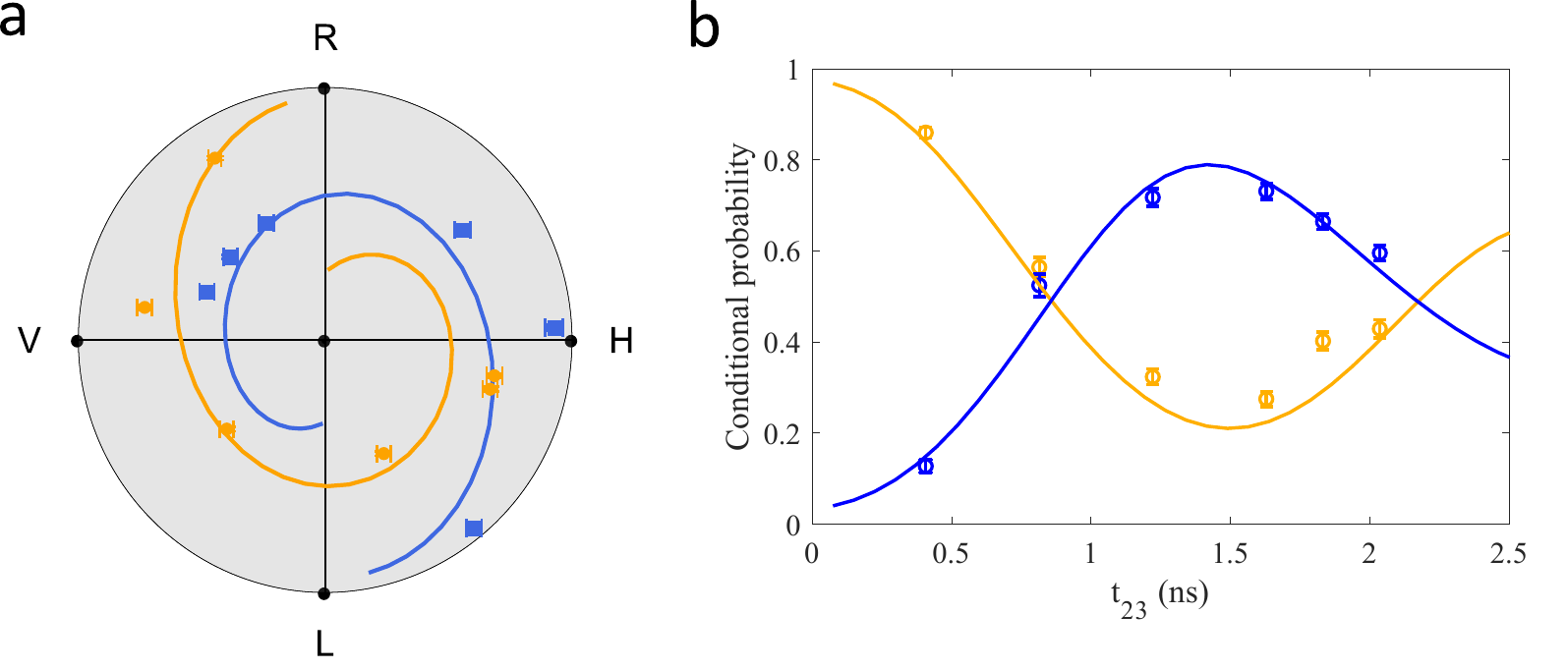}
	\caption{  \textbf{Process fidelity measurement} {a. Measured (symbols) and simulation (solid line) Bloch vector  of the second {emitted} photon {as a function of $t_{23}$. Note that the Bloch vector is measured in the three polarization bases and is found to evolve in a plane that we represent here.  $t_{23}=0$ corresponds to the Bloch vector being at surface of the Bloch sphere. {The distance from the center of the plot indicates a lower bound on the polarization purity, which only lacks contribution from the negligible out-of-plane ($D/A$) component.}}
 In orange (blue) the last photon is measured with $R_3$ ($L_3$) polarization. b. Associated conditional probabilities $P(R_2|R_3)$ (orange) and $P(R_2|L_3)$ (blue) as a function of the $t_{23}$ delay. Modeling these measurements allows to extract the fidelity of a single process step of the cluster state generation scheme (see text). }}

\label{Fig_3}
\end{figure}

    While these two- and three-partite fidelity lower bounds, $F_{s,p}$ and $F_{s,2p}$,  
    are directly deduced from experimental measurements, note that by construction they are not tight {because we overestimate the negative influence of the spin dephasing in their derivations.}
    {To get a better estimate of these fidelities  and a better understanding of the physics at play in the system,} {  we now precisely map photon $\#$2 polarisation to the spin state as a function of the spin readout delay $t_{23}$. To do so, we measure the Bloch  vector characterizing the polarization state of photon $\#$2 as a function of the delay $t_{23}$. This corresponds to 3-photon correlation measurements, detecting photon $\#$1 along the $R$ polarization, photon $\#$3 along $R$ or $L$ polarization and photon $\#$2  along the three polarization axes of the Bloch sphere. 
   The measured Bloch  vector is observed to rotate as a function of $t_{23}$ in a plane of the Bloch sphere that we represent in Fig.\ref{Fig_3}.a. The rotation of photon $\#$2 Bloch vector is correlated with the precession of the spin during $t_{23}$ and  its  length,  reflecting a loss of polarization purity, to the spin decoherence.     {Fig.\ref{Fig_3}.b.}  presents  the parity 
    measurement along $z$ between photon $\#$2 and
    {the spin (or equivalently between photons $\#$2 and $\#$3) as a function of $t_{23}$, measuring the spin (photon $\#$3) in either the $|\uparrow\rangle$ ($\ket{\sigpp}$) state 
    or $|\downarrow\rangle$ ($\ket{\sigmm}$) state. }}

    \noindent These measurements are simulated using a numerical model that solves the master equation using the quantum regression theorem in the Schrödinger picture, conditional evolution and photon number decomposition{~\cite{Wein2020}}.
    
    By fitting our experimental data {(lines in Fig.\ref{Fig_3})}, we 
    {estimate} the relevant parameters of the spin-photon interface (see supplementary) and simulate one entire step of the Lindner and Rudolph protocol, i.e. the emission process equivalent to a spin-photon control-NOT gate~\cite{lindner_proposal_2009}  followed by a $\pi/2$ spin rotation. We can thus characterize the real emission and spin rotation process $C(\rho)$, allowing us not only to provide a simulated density matrix of the spin-photon entanglement, but also to extrapolate the spin and $k$ photon density matrix obtained by $k$ repetitions of this step, $\rho^{(k)}= C^{(k)}(\rho_s)$ (with $\rho_s$ the  spin state after initialisation). We finally deduce the following simulated fidelities of
    {F=\{80\%, 63\%, 50\%, 41\%\},} for the  fidelity to the $n$-partite linear cluster state with one spin and $k=n-1= \{1,2, 3, 4\}$  photon(s) respectively. From our simulations, we find that the exact choice of linear excitation polarization can impact the ideal $t_{12}$ time, and hence {slightly} change the fidelity, since it imprints a phase on the excited hole spin state, an effect that will be the subject of a future work.

We now qualify the photonic nodes of our cluster states in terms of quantum purity and brightness. We measure a second order intensity correlation  of $g^{(2)}(0)=4\pm 0.2\%$ for $B=40$~{mT} evidencing the single-photon nature of the quantum light. The photon indistinguishability is measured by performing the Hong-Ou-Mandel interference of {successive} single photons emitted 
$12\ $ns {apart}, without spin initialization and projecting the single photon onto a linear polarization. At zero magnetic field, a single photon mean-wavepacket overlap of $M=93.0\pm{0.5}\%$ is measured, a value that decreases as the applied magnetic field increases (Fig.\ref{Fig_4}). The reduction is particularly significant when the magnetic field-induced splitting of the trion state exceeds the radiative decay of the photon. This is evidenced by the observation of beatings in the zero delay peaks of the Hong-Ou-Mandel interference shown in the insert of Fig.\ref{Fig_4}. At $B=40$~mT, $M$ remains as high as $88\pm{0.5}\%$. Such value could be further increased using higher acceleration of spontaneous emission leading to a larger intrinsic spectral linewidth for each transition.

\begin{figure}
	\includegraphics[width=0.8\linewidth]{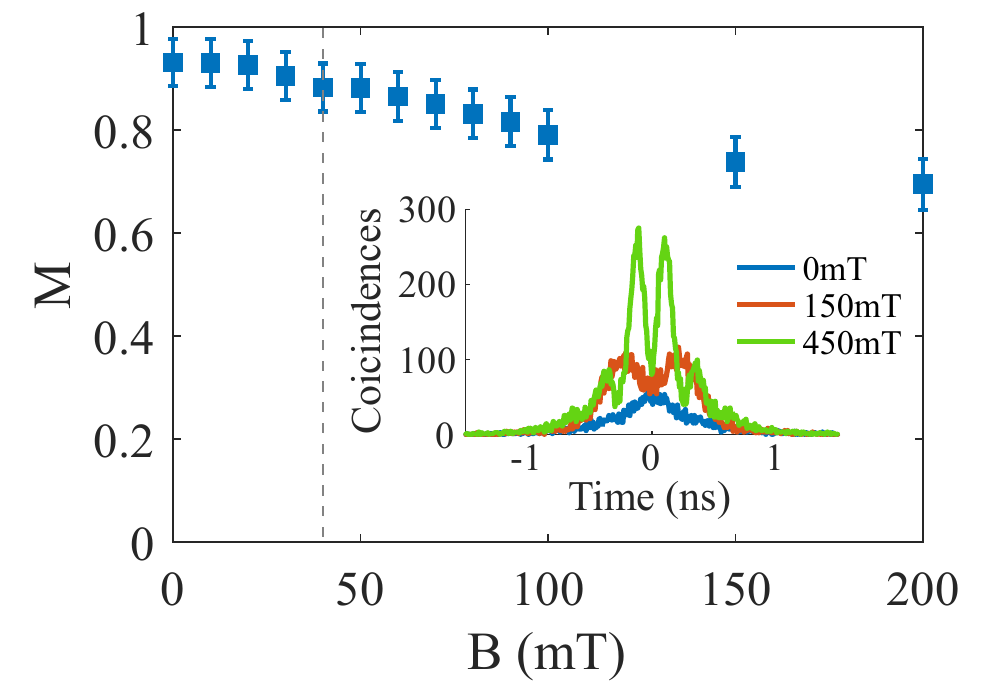}
	\caption{\textbf{
	{Photon indistinguishability}} Measured wavepacket overlap $M$ as a function of the applied Voigt magnetic field. The indistinguishability of the photons at the nodes of the cluster state is above  88\% ($B=40\ $mT - dotted line). Insert: zero delay peak of the Hong-Ou-Mandel measurement for three magnetic field amplitudes. 
	} \label{Fig_4}
\end{figure}

\begin{table}[h!]
\centering
\begin{tabular}{|p{1.9cm}|p{1.3cm}|p{1.9cm}|p{2.1cm}|} 
 \hline
 \fixtable{1.9cm}{Rate (MHz)}  & \fixtable{1.3cm}{Single-photons (n=1)} &   \fixtable{1.9cm}{Heralded spin-photon (n=2)} &   \fixtable{2.1cm}{Heralded spin-photon-photon (n=3)}  \\
 \hline
 \fixtable{1.9cm}{At the first lens} & \fixtable{1.3cm}{15} & \fixtable{1.9cm}{2.8} & \fixtable{2.1cm}{0.52} \\  \hline
\fixtable{1.9cm}{In fiber} & \fixtable{1.3cm}{6.5} & \fixtable{1.9cm}{0.52} & \fixtable{2.1cm}{0.041} \\ \hline
\fixtable{1.9cm}{n-correlation rate} & \fixtable{1.3cm}{4.5} & \fixtable{1.9cm}{0.25}  & \fixtable{2.1cm}{0.014} \\  \hline
\end{tabular}
\caption{\textbf{Entanglement generation rates}. {Single-photon (n=1), heralded spin-photon (n=2), and heralded spin-photon-photon (n=3) generation rates at the output of the device (at the first lens) or coupled to a single-mode fiber. The last line corresponds to  n-particle correlation rates that would be measured in a correlation setup without passive demultiplexing of the photons (see text).} }
\label{table1}
\end{table}	

 We finally discuss the generation rates for single photons, heralded spin-photon entanglement and spin-photon-photon entanglement. We have measured the efficiency of the three sub-parts of our setup: (i) the collection setup of efficiency $\eta_C=0.43$, (ii) the single-photon tomography setup of efficiency $\eta_T=0.69$ and (iii) the three path passive demultiplexer of single-photon efficiency $\eta_D=0.18$ (see supplementary). This yields a total setup efficiency of $\eta_s=\eta_C \eta_T\eta_D= 0.053$, which considering the measured unpolarized single-photon rate of $800$~kHz for an excitation rate of  $f=81$~MHz corresponds to a probability to collect a photon at the first lens (or  first lens brightness) of  $B_{FL}=18.6 \%$. Table \ref{table1} then summarizes the  single-photon (n=1), heralded spin-photon (n=2) and heralded spin-photon-photon (n=3) entanglement rate at the first lens $f\ B_{FL}^n$ and in a fiber $f\ B_{FL}^n \eta_C^n$. To the best of our knowledge, the reported 3-partite entanglement rates of 0.52 MHz (0.041 MHz) at the output of the device (in a fiber),  represent  record values for a solid-state spin~\cite{CoganClusterHole2021}. To compare with the most efficient demonstration so far of multiphoton entanglement obtained with a single atom~\cite{RempeCluster2022}, the last row of Table \ref{table1} shows the multi-photon rate obtained at the end of the tomography setup $f\ B_{FL}^n \eta_C^n \eta_T^n$ in a configuration corresponding to a correlation setup without the passive demultiplexing used here. The two- and three- partite entanglement rates obtained with our QD-cavity system exceeds by roughly  3 and 2 orders of magnitude respectively the recent report. This significant increase is obtained thanks to the high repetition rate $f$ allowed by the QD system that makes up for a first lens brightness $B_{FL}$ more than twice smaller than in {reference}~\cite{RempeCluster2022}.

In conclusion, we have reported on the  efficient generation of linear cluster states with a single solid-state spin and indistinguishable photons up to three particles. The achieved rates exceed by more than two orders of magnitude  the previous state of the art, including the recent atom based approaches~\cite{RempeCluster2022,JWP2021cluster}. This is obtained with semiconductor devices that have already proven reproducible  performances as single-photon emitters~\cite{Ollivier2020} and with a simple experimental configuration. Indeed, the use of acoustic-phonon assisted excitation relies on simple spectral filtering  of the excitation laser and achieves high single photon rate stability~\cite{thomas_bright_2021}. Moreover, we use a  magnetic field of only tens of mT that can be easily implemented with a permanent magnet in {a} standard closed-cycle cryostat. 
We also note that there is {a} large room for improvement{s} to obtain longer cluster states from a QD-in a cavity device at higher rates and indistinguishability. In the present demonstration, the entanglement generation rates are limited by  the occupation probability of the negatively charged QD  that is around  $50\ \%$~\cite{hilaire_deterministic_2020}. In future devices, such {a} limitation can be overcome by having an electrical control of the charge state  during the in-situ lithography process to tune the cavity corresponding to the  energy of maximum occupancy of the trion state~\cite{Dousse2008}. Higher Purcell factor and fine tuning of the cavity geometry should also allow  higher photon collection efficiency~\cite{Gregersen2020}  and also lead to  higher indistinguishability of the photon by reducing the effect of the magnetic field on the trion transition energy. Finally, spectacular progresses in nuclear spin control in QDs has recently allowed spin coherence in the hundred of microsecond time scale~\cite{Zaporski2022}.  All these features put the QD-based technology in an excellent position to provide highly identical long trains of photons in a linear cluster state, an important milestone for scaling-up optical quantum technologies. 

\vspace{0.5cm}

\noindent \textbf{Acknowledgements.} This work was partially supported by the the IAD-ANR support ASTRID program Projet ANR-18-ASTR-0024 LIGHT, the QuantERA ERA-NET Cofund in Quantum Technologies project HIPHOP, the European Union's Horizon 2020 FET OPEN project QLUSTER (Grant ID 862035), the European Union's Horizon 2020 Research and Innovation Programme QUDOT-TECH under the Marie Sklodowska-Curie Grant Agreement No. 861097 and the French RENATECH network, a public grant overseen by the French National Research Agency (ANR) as part of the "Investissements d'Avenir" programme (Labex NanoSaclay, reference: ANR-10-LABX-0035).  N.C. acknowledges support from the Paris Ile-de-France R\'egion in the framework of DIM SIRTEQ. S.C.W. acknowledges support from the Foundational Questions Institute Fund (Grant No. FQXi-IAF19-01). {S.E.E. acknowledges supported from the  from the NSF (Grant No. 1741656)}
\vspace{0.5cm}

\noindent \textbf{Methods}

\noindent \textbf {Sample and experimental procedure. }
The QD-cavity devices are fabricated from a planar sample embedding InGaAs quantum dots at the center of a $\lambda$-cavity composed by two distributed Bragg reflectors  made of GaAs/Al$_{0.9}$Ga$_{0.1}$As $\lambda/4$ layers with 28 (14) pairs {for} the bottom (top) reflector.
The vertical structure includes a p-i-n junction and a 20-nm thick tunneling barrier
of Al$_{0.1}$Ga$_{0.9}$As, positioned 10nm above the QD layer. The cavity is connected to planar mesa where the top electrical contact is defined to apply a bias voltage. 
The sample is placed in a cryostat operating at 5K where two magnetic field coils 
{generate} up to 500 mT of in-plane tunable magnetic field.
The excitation is provided by a Ti:Sa laser, emitting 3 ps pulses with a repetition rate of 81~MHz. The pulses are spectrally filtered to obtain ~15 ps pulse, divided and recombined to generate train of three pulses every 12~ns using beamsplitters and manually adjustable, free-space delays. The detailed setup description and efficiency are provided in Supplementary materials. 

\noindent \textbf {Theory.} 
{Experimental data are theoretically accounted for using master equation simulations based on a four level system (two ground spin states and two excited trion states). We account for the (Purcell-enhanced) spontaneous emission with Lindblad operators and we simulate the hyperfine interaction through an Overhauser field with an isotropic Gaussian distribution. We use the Heisenberg input-output relations to simulate the two- and three- photon correlations measurements. From these simulations, we extract a realistic process map of one repetition step of the Lindner and Rudolph protocol~\cite{lindner_proposal_2009} that we use to estimate the spin-$n$-photon fidelity to a linear cluster experimentally accessible with our device. More information about the theoretical model are available in the Supplementary information.}

\noindent \textbf{Authors contributions.} 
N.C.: experimental investigation, data analysis, methodology, visualization, writing D.F.:experimental investigation, data analysis, methodology, visualization, writing,  N.B.:data analysis, methodology, formal analysis, visualization, writing, supervision, S.C.W.: conceptualization, formal analysis, writing, P.H.: conceptualization, formal analysis, writing,  R.F.:conceptualization, formal analysis, M.G.: formal analysis, B. G.: formal analysis, N. S.: nano-processing, A. L.: sample growth, M.M.: sample growth, I.S.: nano-processing, A.H. nano-processing, S.E.E.: conceptualization, formal analysis, A.A.: formal analysis, O.K.:data analysis, methodology, L.L.: sample design, methodology, data analysis, formal analysis, P.S.: nano-processing, data analysis, methodology, visualization, writing, supervision, funding acquisition.


\textbf{Data and materials availability:}
All data acquired and used in this work is property of the Centre for Nanoscience and Nanotechnology and is available upon reasonable request to pascale.senellart-mardon@c2n.upsaclay.fr or nathan.coste@c2n.upsaclay.fr.\\

%



\clearpage

\title{Supplementary material: High-rate entanglement between a semiconductor spin and indistinguishable photons }

\maketitle


\section{Extended data}

\subsection{Photon $\#$2 parity measurement in the linear basis.}

In the main text, we presented the evolution of the Bloch vector for photon $\#$2 conditioned on the photon $\#$3 being measured in either R or L polarisation (Fig. 3.a.) and the corresponding parity measurements in the R/L basis for photon $\#$2. We present below complementary data, showing the parity measurement in the H/V basis as a function of $t_{23}$. 

\begin{figure}[h]
	\includegraphics[width=0.8\linewidth]{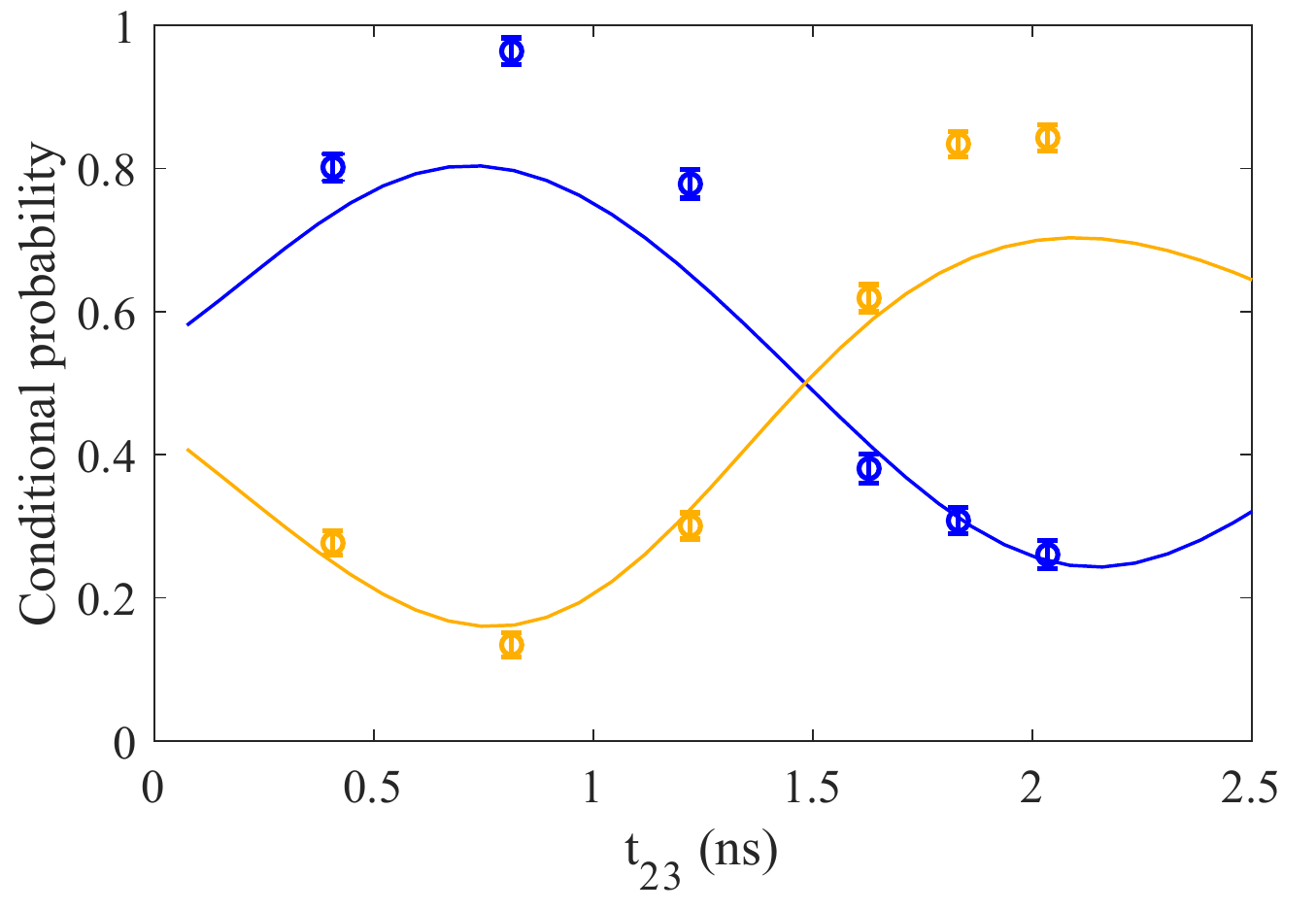}
	\caption{Symbols: conditional probability to measure photon $\#$2 in H as a function of $t_{23}$ when measuring photon $\#$3 in R (orange) or L (blue). Lines: fit to the data with the same parameters as in Fig.3. of the main text.  \label{Fig: Conditional probability HV2}}
\end{figure}







 {\subsection{Single photon purity and indistinguishability measurements}}

\begin{figure}[h]
	\includegraphics[width=0.8\linewidth]{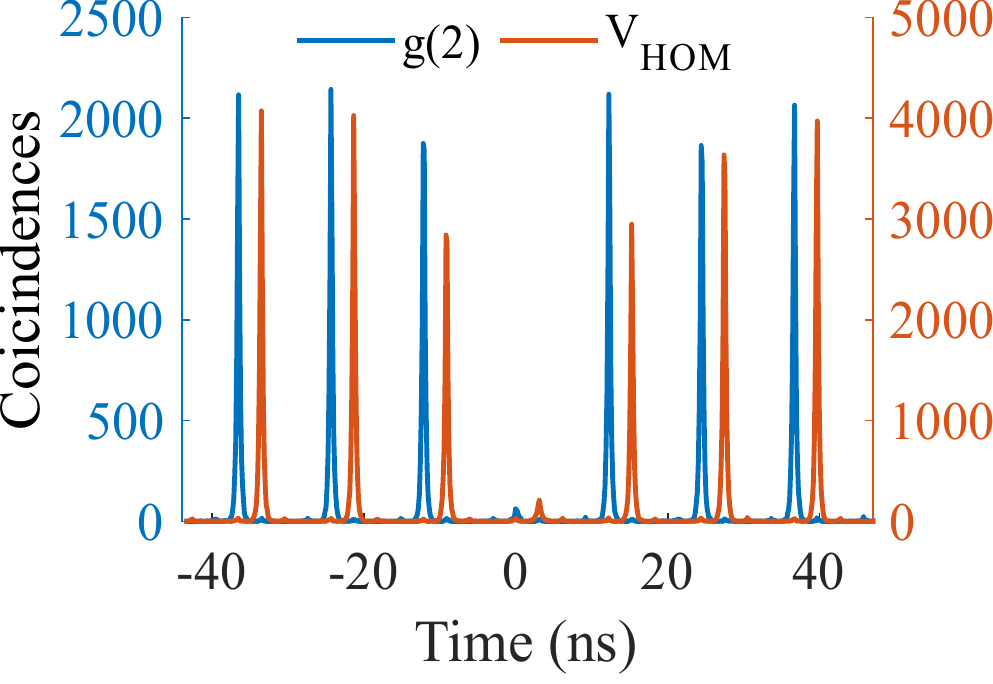}
	\caption{Example of second-order intensity correlation measurements giving  access the source single-photon purity (blue) and single-photon indistinguishability (red) }\label{Fig: g2HOM}
	
\end{figure}

Figure \ref{Fig: g2HOM} presents an example of second-order intensity correlation measurements giving  access the source single-photon purity (blue) and single-photon indistinguishability (red). The red data are shifted by +3 ns for clarity purpose. Both curves are obtained at $B=0$. The blue curves demonstrates  $g^{(2)}(0) = 4{\pm 0.2\%}$ and the red curve a Hong-Ou-Mandel visibility of  $V_\text{HOM} = {88\pm 0.5}\%$. Considering the distinguishable nature of the additional photons leading to a non-zero $g^{(2)}(0)$, we deduce a single photon mean-wavepacket overlap of $M_s={93}\pm0.5\%$~\cite{THomas_HOM_2021}. 

\clearpage
\begin{figure*}[t]
	\includegraphics[width=0.7\linewidth]{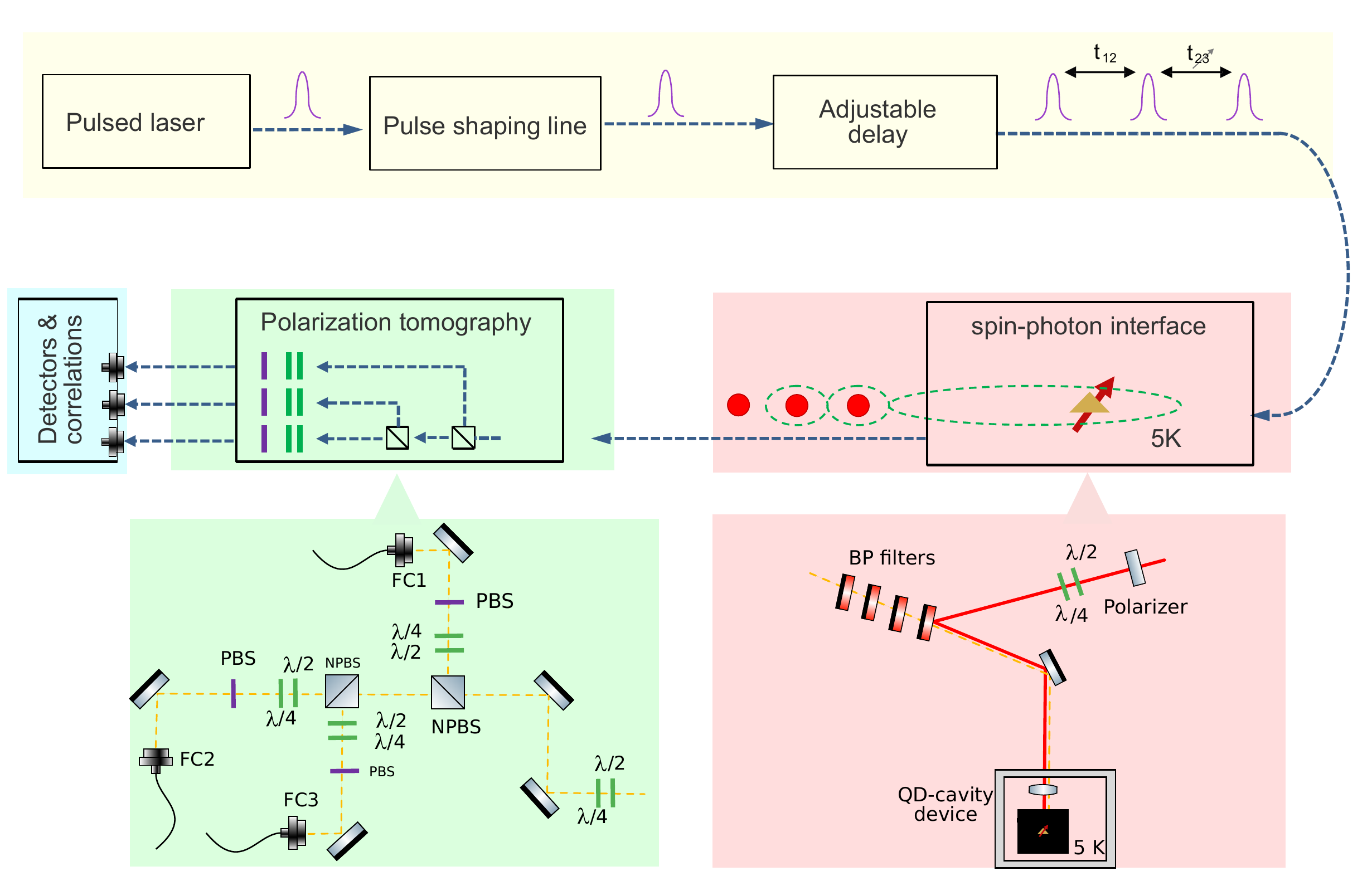}
	\caption{ { \textbf{Schematics of the experimental setup.} The experimental setup is split in 4 parts: (i) the excitation pulse preparation (yellow) (ii) the spin-photon interface and photon collection (pink)  (iii) the tomography setup (green) using a passive  photon multiplexing with non-polarizing beam-splitters (NPBS), quarter ($\lambda/4$) and half ($\lambda/2$) waveplates, polarizing beam-splitters (PBS) and fiber couplers (FC1, FC2, FC3) (iv) the detection and correlation module (blue). 
	} \label{Fig: SketchSetup}}
\end{figure*}

\section{Experimental procedure}

The quantum dot is excited with the train of linearly polarized laser pulses {(Fig. \ref{Fig: SketchSetup})} leading to a QD unpolarized  single-photon emission. $t_{12}$ is chosen so that for $2t_{12}$ the spin has   undergone half a rotation and is most probably in the $|\downarrow \rangle$ state.  Experimentally,  this corresponds to a maximum antibunching between the first and the second emitted photon measured in the R/L basis. {The photons emitted by the QD} are collected in a confocal microscope configuration. The {spectral} detuning between the excitation laser and the emitted photons by the quantum dot allows to use high transmission (92\%), narrow band-pass filters (0.8nm bandwidth) to separate the excitation pulses from the collected photons {(transmitted)}. 

\noindent Considering the smallest temporal separation of the QD emitted photons of $\approx 800$~ps and the dead-time of tens of nanoseconds of the nanowire superconducting detector, a passive demultiplexing scheme and three detectors are used to perform correlation measurements. The chain of photons is sent through a 70:30 non-polarizing beamsplitter (NPBS), followed by a 50:50 NPBS, sending each photon with roughly equal probability to one of three tomography paths. Each path comprises a set of {motorized} quarter and half waveplates and a polarizing beamsplitter PBS used as a polarizer to perform polarisation tomography. Detection events are recorded and analyzed by a Swabian Time Tagger 20 correlator.


 All arrival times of the photons are recorded and digitally shifted so that a $(t_{12},t_{23})$ event is detected as a 3-photon-coincidence. For each delay {$t_{23}$} between the second and third pulses, such 3-photon-coincidences are recorded for 12 settings of waveplates, corresponding to the detection of photons in R polarisation for the detection path corresponding to photon $\#$1, H,V,D,A,R or L for photon $\#$2, and R or L for photon $\#$3.
 
 \clearpage

\subsection{Loss budget} 

We measured all component transmissions in order to provide the three efficiencies mentioned in the main text: (i) the collection setup of efficiency $\eta_C$, (ii) the single-photon tomography setup of efficiency $\eta_T$ and (iii) the three path passive demultiplexer of single-photon efficiency $\eta_D$. 

\begin{table}[h]
    \centering
\begin{tabular}{ |c|c|} 
\hline
 & Transmission \\
\hline
Lens and cryostat window
 & 0.89 \\ 
\hline
Excitation waveplates and mirrors & 0.92 \\ 
\hline
4 Band pass filters & 0.70 \\ 
\hline
Fiber coupling & 0.75 \\ 
\hline
{\bf Collection setup efficiency $\eta_{C}$} & 0.43 \\
\hline
\end{tabular}
    \caption{
    {Single}-photon collection efficiency.}
    \label{tab:my_label}
\end{table}

\begin{table}[h]
    \centering
\begin{tabular}{ |c|c|} 
\hline
 & Transmission \\
\hline
2 waveplates  and polarizing beam-splitter & 0.86 \\ 
Fiber transmission & 0.9 \\ 
Detector efficiency & 0.90 \\\hline
{\bf Tomography setup efficiency $\eta_{T}$ }& 0.69 \\
\hline
\end{tabular}
    \caption{
    {Single}-photon tomography setup efficiency.}
    \label{tab:my_label}
\end{table}

\begin{table}[h]
    \centering
    \begin{tabular}{ |c|c|}
\hline
 & Transmission \\
\hline
Non polarizing beam splitter 1 & 0.63 \\ 
Non polarizing beam splitter 2 & 0.41 \\ 
Fiber connector & 0.7 \\\hline
{\bf Demultiplexing setup efficiency $\eta_{D}$} & 0.18 \\
\hline
\end{tabular}
    \caption{Passive three-photon demultiplexing efficiency.}
    \label{tab:my_label}
\end{table}

\noindent {Note that the implementation of our tomography setup is not optimized, making in particular use of non polarizing beamps-plitters. Moreover, the  passive demultiplexing introduces  additional losses that could be avoided with fast switching devices.} 

\clearpage

\section{Derivation of the spin-photon-photon fidelity lower bound}

The objective of this section is to demonstrate that we have a lower bound $F_{s,2p}$ on the fidelity to $\ket{\psi_3}$ for the generated the spin-photon-photon. 
$F_{s,2p}$  is derived from  the experimentally demonstrated spin-photon state fidelity lower bound $F_{s,p}$, using two hypotheses which are experimentally validated:

\begin{hyp}
    {Would the system emit two photons with a delay $\Delta t \to 0$, the two emitted photons would always be measured} to have the same circular polarisation (either both $\ket{R}$ or both $\ket{L}$).
    \label{hyp_zz}
\end{hyp}
{This hypothesis is consistent with the fact that the polarisation selection rules are very well preserved by the LA-phonon-assisted scheme, as recently demonstrated~\cite{Coste-precession-2022} and at the low value of magnetic field used for the protocol. Moreover, the Purcell accelerated  emission lifetime of the trion is short compared to the hole spin dynamics, resulting in a negligible  hyperfine interaction with the nuclear spins in the excited state~\cite{Urbaszek2013}. 
Note that this is experimentally consistent  with the theoretical interpolation of the parity measurements of Fig.3.b. Indeed, these  data show that starting from a spin state in the $\ket{+_s}$  state at $t_2$, the parity curves show that photons $\#2$ and $\#3$ are measured with the same R or L polarisation for  $t_{23}=0$. This is also visualized in the Bloch sphere in Fig. 3.a. where the conditional Bloch vector of photon $\#2$ points toward the R (L) pole depending on photon $\#3$ being measured  in the R (L) basis. 
}

\begin{hyp}
At the beginning of an experiment, the spin is in a maximally-mixed state { i.e. is found in all orientations with equal probabilities}.
    \label{hyp_mixed}
\end{hyp}
{This is experimentally verified by the observation of a residual polarisation of the emitted photons of only a few percent after the first excitation pulse}.

\vspace{0.5cm}
\noindent Starting from a general unknown emission process $C(\rho)$, we 
show that these two hypotheses and our experiments add constraints  
{to the generic} process, such that after emission of photon $\#3$  a lower bound for the fidelity $F_{s,2p}$ to the state $\ket{\psi_3}$  can be derived and is given by Eq.~\eqref{eq_f_s_2p}.

\subsection{Emission of photon $\#3$}

After the second photon emission and given the  spin-photon fidelity lower bound $F_{s,p}$, the {spin-photon} system is described by a density matrix 
\begin{align}
    \rho_{s,p}=F_{s,p} & \rho_{2} + (1 - F_{s,p})\rho_e,\\
&\rho_{2} = \ketbra{\psi_2}{\psi_2} ,
&&\ket{\psi_2} = \frac{1}{\sqrt{2}} (\ket{\uparrow, R_2} + \ket{\downarrow, L_2})\nonumber
\end{align}
with $\rho_e$ an unknown density matrix and  $\rho_{2}$ the target ideal state.

We compare the real emission process $C(\rho)$ to the ideal emission process $C^{\rm ideal}(\rho) = K_0 \rho K_0^{\dag}$ 
where $K_{0}$ is an isometry $K_{0}=|0_s 0_{ph}\rangle\langle 0_s| +|1_s 1_{ph}\rangle\langle 1_s|$. $K_{0}$ is written in the computational qubit basis i.e. where $\ket{\uparrow}$ ($\ket{\downarrow}$) is assigned $\ket{0_s}$ ($\ket{1_s}$) and $\ket{R}$ ($\ket{L}$) is assigned $\ket{0_{ph}}$ ($\ket{1_{ph}}$).
The newly generated spin-photon-photon state $\rho_{s,2p} = C(\rho_{s,p})$ is then compared to the ideal pure target state $\rho_{3} = |\psi_{3}\rangle\langle\psi_{3}|$ with $|\psi_{3}\rangle = K_0 |\psi_2\rangle$.
The fidelity $F$ of the final state {$\rho_{s,2p}$} to $\rho_{3}$ is {thus lower bounded}
\begin{align}
    F & = \langle \psi_{3}|\rho_{s,2p}|\psi_{3}\rangle \\
    & \geq F_{s,p} \langle \psi_{3}|C(\rho_{2})|\psi_{3}\rangle \\
    & = F_{s,p} \langle \psi_2| K_0^\dag C(\rho_{2}) K_0|\psi_2\rangle.
\end{align}

A lower bound for the spin-photon-photon state fidelity {$F$} can be found by deriving a lower bound for the term $A=\langle \psi_2| K_0^\dag C(\rho_{2}) K_0|\psi_2\rangle$, and thus acquiring knowledge on the emission process $C(\rho)$.

\subsection{Realistic emission process}
The most general emission process can be written as
\begin{equation}
    C(\rho) = \sum_i p_i K_i \rho K_i^{\dag}
\end{equation}
where $\sum_{i} p_{i} = 1$ and the emission operators $K_i$ have the form
\begin{equation}
    K_i = |\psi^{(i)}_0\rangle \langle0_s| + |\psi_1^{(i)}\rangle \langle1_s|,
\end{equation}
where the $|\psi^{(i)}_j\rangle$ are arbitrary spin-photon states.

\noindent  Hypothesis~\ref{hyp_zz} implies that
two successively emitted photons have the same polarisation either $\ket{0_{ph}}=\ket{R}$ or $\ket{1_{ph}}=\ket{L}$.
Consequently,
$|\psi^{(i)}_j\rangle  = |\xi^{(i)}_{s,j}\rangle |m^{(i)}_{ph,j}\rangle$ with $|m^{(i)}_{ph,j}\rangle = |0_{ph}\rangle$ or $|1_{ph}\rangle${, and where $\ket{\xi^{(i)}_{s,j}}$ is an arbitrary spin state}.
So this adds a constraint on the emission operators which should have the form:
\begin{equation}
    K_i = |\xi^{(i)}_{s,0}\rangle |m^{(i)}_{ph,0}\rangle\langle 0_s| + |\xi^{(i)}_{s,1}\rangle |m^{(i)}_{ph,1}\rangle\langle 1_s|.
\end{equation}

\noindent By applying twice $K_i$ onto $|j_s\rangle\langle j_s|$, and only considering the photonic subsystem density matrix (i.e. using a partial trace on the spin subspace), we obtain
\begin{align}
     {\rm Tr}_{\rm spin}[K_i K_i\ketbra{j_s}{j_s} K_i^\dag K_i^\dag]& =|\langle 0_s|\xi^{(i)}_{s,j}\rangle|^2 |m^{(i)}_{ph,0}, m^{(i)}_{ph,j}\rangle \langle m^{(i)}_{ph,0}, m^{(i)}_{ph,j}| \nonumber \\ & + |\langle 1_s|\xi^{(i)}_{s,j}\rangle|^2 |m^{(i)}_{ph,1}, m^{(i)}_{ph,j}\rangle \langle m^{(i)}_{ph,1}, m^{(i)}_{ph,i}|
\end{align}

From Hypothesis~\ref{hyp_zz}, it follows that:
\begin{itemize}
    \item either $\langle 1_{s} | \xi^{(i)}_{s,j}\rangle = 0$ or $\langle 0_{s} | \xi^{(i)}_{s,j}\rangle = 0$. In that case, the emission operators belong to
     \begin{align}
        S_{0} = \{K_{0,\phi} =|0_s,0_{ph}\rangle\langle 0_s| + e^{i \phi} |1_s,1_{ph}\rangle\langle 1_s|, \phi \in \left[0, 2\pi\right[\} &
        \intertext{or}
         S_{1} = \{K_{1,\phi}=|0_s,1_{ph}\rangle\langle 0_s| + e^{i \phi} |1_s,0_{ph}\rangle\langle 1_s|, \phi \in \left[0, 2\pi\right[\}.
    \end{align}
    The two terms in the emission operator are $\ketbra{0_s,0_{ph}}{0_s}$ and $\ketbra{1_s,1_{ph}}{1_s}$ (or $\ketbra{0_s,1_{ph}}{0_s}$ and $\ketbra{1_s,0_{ph}}{1_s}$). We have also kept a phase difference $\phi$ between these two terms.
    \item or we have a mixed superposition of $\ket{m_{ph,0}^{(i)},m_{ph,j}^{(i)}}$ and $\ket{m_{ph,1}^{(i)},m_{ph,j}^{(i)}}$. Verifying Hypothesis~\ref{hyp_zz} implies $m_{ph,0}=m_{ph,1}$. In that case the resulting emission operators have the form
    $S_2 = \{K_{\xi, 2} = (|\xi_{s, 0}\rangle \langle 0_s| + |\xi_{s, 1}\rangle \langle 1_s|) \otimes |0_{ph}\rangle \}$: $\ket{0_{ph}}$ emitted photons only, 
    or 
    $S_3 = \{K_{\xi, 3} = (|\xi_{s, 0}\rangle \langle 0_s| + |\xi_{s, 1}\rangle \langle 1_s|)  \otimes |1_{ph}\rangle \}$: $\ket{1_{ph}}$ emitted photons only.
\end{itemize}

We can thus divide the general process in these four subcategories:
\begin{align}
    C(\rho) & = \sum_{j=0}^{3} p'_j C_j(\rho) & & \sum_{j=0}^{3} p'_{j} =1\\
    \text{with \;} C_j(\rho) & = \sum_{\substack{k\\K_k \in S_j}} p_{k,j} K_k \rho K_k^{\dag} & & \sum_{k} p_{k,j} =1
\end{align}

Because $C_{i} \circ C_{j} (\rho)$ produce two photons with the same circular polarisation only when $i = j$, Hypothesis~\ref{hyp_zz} implies that $C(\rho)$ is actually either $C_{0}(\rho)$, $C_{1}(\rho)$, $C_{2}(\rho)$, or $C_{3}(\rho)$.
Besides, Hypothesis~\ref{hyp_mixed} is incompatible with $C(\rho) = C_2(\rho)$ or  $C(\rho) = C_3(\rho)$ which both only emit polarized photons.
Therefore, 
$C(\rho) = C_{0}(\rho)$ or $C_{1}(\rho)$.
Note that these two cases are identical up to a change of notation $\ket{0_s} \leftrightarrow \ket{1_s}$ for the spin state, so we can choose the correct process to be of the form $C_0(\rho)$ without loss of generality.
The process can thus be written
\begin{equation}
    C(\rho) = \int_{\phi} p(\phi) K_{\phi} \rho K_{\phi}^{\dag} \mathrm{d}\phi,
    \label{eq_correct_c}
\end{equation}
with $K_{\phi}=|0_s 0_{ph}\rangle\langle 0_s| + e^{i \phi} |1_s 1_{ph}\rangle\langle 1_s| $ (given that these operators only differ by a phase, we use an integral sum).
This process accounts for phase jitter in the emission process.

\subsection{Lower bound for $A$}

Given Eq.~\eqref{eq_correct_c}, we can already simplify $A = \langle \psi_2| K_0^\dag C(\rho_{2}) K_0|\psi_2\rangle$ given that:
\begin{align}
    A & = \int_{\phi} p(\phi)|\langle \psi_2| K_0^{\dag} K_{\phi}|\psi_2\rangle|^2 \mathrm{d}\phi
    \intertext{where $K_0^{\dag} K_{\phi}$ is a $2 \times 2$ matrix}
    K_0^{\dag} K_{\phi} & = \begin{pmatrix} 1 & 0 \\ 0 & e^{i \phi}    \end{pmatrix}.
\end{align}
We find
\begin{equation}
        A = \int_{\phi} p(\phi)|1 + e^{i \phi}|^2 \mathrm{d}\phi.
\end{equation}
We now show that $P(V_2|\uparrow)$ (at $t_{12} = t_{23}$ in the 3-photon correlation experiment) is a lower bound of $A$.

In the following, we consider a perfect spin control, initialisation and measurement, which in reality are imperfect and will cast all the imperfections on the emission process. By doing so, we derive more simply a lower bound for the emission process which still holds in the  real case of an imperfect spin manipulation. 

The state before photon $\#2$ emission is $\rho_+ = |+_s\rangle\langle+_s|$, with $|+_s\rangle = (|0_s\rangle + |1_s\rangle)/\sqrt{2}$ and immediately after emission we have:
\begin{equation}
    C(\rho_{+})=\int_{\phi} p(\phi) K_{\phi} \rho_+ K_{\phi}^{\dag} \mathrm{d}\phi
\end{equation}

The operation of waiting for a $\pi/2$ rotation and measuring the spin corresponds to a spin measurement in the $X$ basis at the time before the spin rotation. Thus, the detection of photon $\#3$ with $\ket{0_{ph}}\leftrightarrow \ket{R}$ or $\ket{1_{ph}}\leftrightarrow \ket{L}$ polarisation (corresponding respectively to a spin $\ket{\uparrow}$ or $\ket{\downarrow}$ measurement at $t_3$) corresponds effectively to a measurement in the $|+_s\rangle$ or $|-_s\rangle$ basis at the time of emission of photon $\#2$.
For a spin measured in $|\uparrow\rangle$ at $t_3$, we find that
\begin{equation}
    P(V_2 |\uparrow) = \int_{\phi} p(\phi)|1 + e^{i \phi}|^2 \mathrm{d}\phi = A.
\end{equation}
Note that here the spin is assumed to be perfectly manipulated. In practice however, $P(V_2 |\uparrow)$ is only a lower bound for $A$: $P(V_2 |\uparrow) \leq A$. Similarly, we can find $P(H_2 |\downarrow) \leq A$, so to account for experimental errors, we take the average of the two: $P' = (1 - s_{X})  / 2$ where $s_{X}= -91.5 \pm 2 \%$ is the projection of photon $\#$2's Bloch sphere vector onto the H/V axis. 

\noindent Consequently, a lower bound for the fidelity of the spin+2 photon state is given by:
\begin{equation}
    F \geq F_{s,2p} = F_{s,p} \frac{1 - s_X}{2}
    \label{eq_f_s_2p}
\end{equation}

\section{Model for the cluster state generation process}
{\subsection{General framework}}
We use a master equation to model the spin-photon interface. The charged quantum dot system is simulated by a four-level system with two ground electron spin states denoted $\ket{\uparrow}$ and  $\ket{\downarrow}$, and two excited hole states  denoted $\ket{\downarrow\uparrow\Uparrow}$ and $\ket{\downarrow\uparrow\Downarrow}$.
The spin Hamiltonian is given by
\begin{equation}
    H_{s} = \frac{\Delta_e}{2} \sigma_y^{(e)} +  \frac{\Delta_h}{2} \sigma_y^{(h)},
\end{equation}
where $\sigma_{y}^{(e)} = i (|\downarrow \rangle \langle\uparrow| - |\uparrow \rangle \langle\downarrow|)$, $\sigma_{y}^{(h)} = i (|\downarrow\uparrow\Downarrow \rangle \langle \downarrow\uparrow\Uparrow| - |\downarrow\uparrow\Uparrow \rangle \langle\downarrow\uparrow\Downarrow|)$, and $\Delta_e$ ($\Delta_h$) is the electron (hole) state splittings.
{The splittings depend} on the applied transverse magnetic field $B$ and the electron (hole) Landé factor $g_e$ ($g_h$)  through $\Delta_e = \mu_B g_e B$ ($\Delta_h = \mu_B g_h B$), where $\mu_B$ the Bohr magneton.

\noindent Due to the solid-state environment, the surrounding nuclear spin bath couples to the electron spin via the hyperfine interaction. We assume that the state of the nuclear spin bath fluctuates on a timescale much slower than the timescale of the quantum dot dynamics, so that the hyperfine interaction is modelled by a mean Overhauser field $\vec B_O$. This interaction is captured by the Hamiltonian
\begin{equation}
    H_O = \frac{1}{2}g_e \mu_B \vec B_O \cdot \vec{\sigma}^{(e)},
\end{equation}
where $\vec \sigma^{(e)} = (\sigma_x^{(e)},\sigma_y^{(e)},\sigma_z^{(e)})$. The Overhauser field is considered to be static for the quantum state evolution, but takes independent Gaussian-random values in each of the three Cartesian coordinates. The quantum dot density matrix $\rho(t)$ is then computed by averaging over the random Overhauser field. This averaging results in an apparent spin-state dephasing that reduces the fidelity of the generated cluster state to the ideal target state. We find that this dephasing model is a more accurate description of the system dynamics compared to a pure-dephasing model, which is consistent with previous studies~\cite{Bechtold2015}. This is especially true when the applied static magnetic field ($\sim40$ mT) is on the same order as the standard deviation of the Overhauser field ($\sim11$ mT), which is the case here. This causes the decoherence dynamics to be poorly described by a single $T_2^*$ dephasing time.

To account for the coupling of the quantum dot optical transitions to the electromagnetic field, we model the quantum dot state evolution using the quantum optical master equation
\begin{equation}
    \dot \rho = \mathcal{L}\rho=- \frac{i}{\hbar} [H_s + H_O, \rho] + \sum_j \mathcal{D}_{A_j}\rho.
\end{equation}
The dissipative Lindblad superoperator is defined by $\mathcal{D}_{A_j}\rho = \mathcal{J}_{A_j}\rho - \frac{1}{2} \left\{ A_j^\dag A_j, \rho \right\}$,
where $A_j$ are the Lindblad collapse operators of the quantum dot system, $\mathcal{J}_{A_j}\rho= A_j \rho A_j^\dag$ is the jump superoperator corresponding to the collapse operator $A_j$, and $\{A,B\} = AB + BA$ is the anti-commutator. We write all linear superoperators with a calligraphic font and assume they act on all operators to their right. The right- ($j=R$) and left- ($j=L$) circular polarized transitions are described by the collapse operators $A_j  = \sqrt{\gamma}\sigma_j$, where $\gamma=1/T_1$ is the decay rate of the trion state, $\sigma_R={\ket{\uparrow}\bra{\downarrow\uparrow\Uparrow}}$ is the $R$-polarized lowering operator, and $\sigma_L={\ket{\downarrow}\bra{\downarrow\uparrow\Downarrow}}$ is the $L$-polarized lowering operator. Note that, although the quantum dot is inside a micropillar cavity, the device is described well by a model assuming a Purcell-enhanced decay rate $\gamma$, since the QD-cavity system operates far into the bad-cavity regime where emission from the quantum dot into the cavity mode is practically irreversible \cite{wein2022photon}.

The dipole coupling to the detected electromagnetic field modes ($a_j^{\rm (out)}$) is modelled using the Heisenberg input-output relations $a_j^{\rm (out)}(t) - a_j^{\rm (in)}(t) = \sqrt{\eta}A_j(t)$ for transition $j\in\{R,L\}$, where $\eta$ incorporates the collection, transmission, and detection efficiency. Since the excitation is filtered in frequency from the output field, we approximate the input field ($a_j^{\rm (in)}$) as being in the vacuum state. This implies an effective dipole proportionality relation $a_j=a_j^{\rm (out)}\propto A_j$ for normally-ordered correlations. Furthermore, the excitation pulses are fast (typically 20 ps) relative to the dynamics of the emission (typically 200 ps) and so we model the pulses as instantaneous linearly-polarized $\pi$-pulse operations $\mathcal{P}_{\theta}\rho=R_{\theta}\rho R_{\theta}^{\dagger}$. The unitary $\pi$-pulse rotation matrix $R_\theta$ of linear polarisation angle $\theta$ is given by
\begin{equation}
    R_{\theta} = e^{-i(\pi/2)\left(\cos\theta\sigma_{y,H}+\sin\theta\sigma_{y,V}\right)},
\end{equation}
where $\sigma_{y,j}=-i(\sigma_j-\sigma^{\dagger}_j)$ for $j\in\{H,V\}$, $\sigma_H=(\sigma_L+\sigma_R)/\sqrt{2}$, and $\sigma_V=-i(\sigma_L-\sigma_R)/\sqrt{2}$.

\subsection{Three-photon polarisation correlations}

A standard approach using quantum regression theory to compute pulsed multi-photon correlation functions requires integrating each detection window over time. Although this is feasible when computing a single solution, it becomes computationally challenging when averaging over many orientations of the Overhauser field or numerically fitting the model to a set of data. For this reason, we take a different approach inspired by quantum trajectories and quantum measurement theory. 

Starting from the master equation $\dot{\rho}=\mathcal{L}\rho$, we perform a photon-number decomposition~\cite{Wein2020} of the source dynamics in powers of the polarisation-dependent jump superoperator $\mathcal{J}_{\vec p}\rho=a_{\vec p}\rho a_{\vec p}^\dagger$ induced by the detection of a photon of polarisation $\vec p$. Here, $a_{\vec p}=\cos\theta a_H+e^{i\phi}\sin\theta a_V$ is the annihilation operator corresponding to the detected polarisation, with $a_H=(a_L+a_R)/\sqrt{2}$ and $a_V=-i(a_L-a_R)/\sqrt{2}$. The general solution to the master equation is then decomposed into $\mathcal{K}(t,t_0)=\sum_{n=0}^\infty\mathcal{K}_{\vec p}^{(n)}(t,t_0)$ where $\mathcal{K}^{(n)}_{\vec p}(t,t_0)$ is the propagator conditioned on the detection of $n$ photons of polarisation $\vec p$ between times $t_0$ and $t$, and where the zero-order term $\mathcal{K}^{(0)}_{\vec p}$ is the general solution to the equation of motion for the state of the system conditioned on no detection: $\dot{\rho}_{\vec p}^{(0)} = (\mathcal{L} - \mathcal{J}_{\vec p})\rho_{\vec p}^{(0)}$. It follows that the dynamics conditioned on the detection of at least one photon between $t_0$ and $t$ is $\mathcal{B}_{\vec p}(t,t_0) = \sum_{n=1}^\infty \mathcal{K}^{(n)}(t,t_0)=\mathcal{K}(t,t_0)-\mathcal{K}^{(0)}_{\vec p}(t,t_0)$, and we name $\mathcal{B}$ the bright propagation superoperator. Hence, the state of the source at time $t$ conditioned on the detection of at least one photon of polarisation $\vec p$ between time $t_0$ and time $t$ is $\rho_{\vec p}(t,t_0)=\mathcal{B}_{\vec p}(t,t_0)\rho(t_0)$. This unnormalized conditional density operator then directly gives the probability of detecting at least one photon between $t_0$ and $t$ through the relation $P_{\vec p}(t,t_0)=\text{Tr}\{\rho_{\vec p}(t,t_0)\}$. 

As a consequence of the quantum regression theorem, this conditional evolution approach can be extended to sequential photon detection patterns. Consider three polarisation measurements described by the set of polarisation vectors ${\bf p}=(\vec p_1, \vec p_2,\vec p_3)$, each performed in the window of time following the pulse arriving at time $t_i$ and before the pulse arriving at time $t_{i+1}$. Then, the conditional state of the quantum dot long after the final photon has been emitted is given by
\begin{equation}
    \rho_{{\bf p}}({\bf t})=\lim_{t\rightarrow\infty}\mathcal{B}_{\vec p_3}(t,t_3)\mathcal{P}_\theta \mathcal{B}_{\vec p_2}(t_3,t_2)\mathcal{P}_\theta \mathcal{B}_{\vec p_1}(t_2,t_1)\mathcal{P}_\theta\rho(t_1),
\end{equation}
for some initial quantum dot state $\rho(t_1)$, which we take to be the mixed state $\rho(t_1)=\left(\ketbra{\uparrow}{\uparrow}+\ketbra{\downarrow}{\downarrow}\right)/2$. The instantaneous pulse superoperator $\mathcal{P}_\theta$ for linear polarisation angle $\theta$ is as defined above.
Thus, we obtain the numerically-exact polarisation-dependent three-photon coincidence probability $P_{{\bf p}}(\bf t)=\text{Tr}\{\rho_{\bf p}(\bf t)\}$ without integrating multi-time correlation functions. For time-independent systems where $\mathcal{K}$ and $\mathcal{K}^{(0)}_{\vec p}$ can be efficiently computed by diagonalizing $\mathcal{L}$ and $\mathcal{L}-\mathcal{J}_{\vec p}$, respectively, this approach can provide orders of magnitude speedup in computational time.

There is a subtle difference between evaluating multi-time intensity correlation functions and using conditional probabilities in the way described above. The former assumes that the detection probability is proportional to the average photon number $\mu=p_1+2p_2+3p_3\cdots$ whereas the latter assumes that the detection probability is proportional to $1-p_0=p_1+p_2+p_3\cdots$, where $p_n$ is the probability that $p_n$ photons arrive at the detector. Both approaches agree in the limit of large losses or when the mode contains at most one photon, because in this limit we have $\mu\simeq p_1 \simeq 1-p_0$. This limit is satisfied for our model, where we use only instantaneous pulses and hence $p_{n\geq 2}=0$ for a given measurement window. It also implies that the choice of global detection efficiency $\eta$ has no impact on the normalized simulated results. However, to use the conditional evolution approach with a spin-photon interface model that predicts multi-photon contributions ($g^{(2)}(0)>0$), it is necessary to choose $\eta$ to be an experimentally realistic value so that the detection regime is accurately captured by the simulation.

The conditional probabilities $P_{{\bf p}}$ are then averaged over the Overhauser field. Due to linearity, averaging at this point is identical to averaging the conditional density matrix itself. With the average detection probabilities, we construct the relevant conditional polarisation probabilities to obtain the polarisation tomography of the second photon, conditioned on the detection of an $R$-polarized photon after the first pulse and either an $R$- or $L$- polarized photon after the third pulse:
\begin{equation}
\begin{aligned}
    P(R_2|R_3) & =&\frac{P_{RRR}}{P_{RRR}+P_{RLR}}\\
    P(R_2|L_3) &=& \frac{P_{RRL}}{P_{RRL}+P_{RLL}}\\
    P(H_2|R_3) &=& \frac{P_{RHR}}{P_{RHR}+P_{RVR}}\\
    P(H_2|L_3) &=& \frac{P_{RHL}}{P_{RHL}+P_{RVL}}
\end{aligned}
\end{equation}
Similarly, we compute the 3 Stokes parameters to obtain the conditional trajectory of the second photon's polarisation in the Poincaré sphere.

Following this approach, and by fixing $T_1=200$ ps$\pm10$ ps, $t_2-t_1=810$ ps$\pm40$ ps, $B=40$ mT, we find that the three-photon correlation data is reproduced well using the parameters $g_e=0.60\pm0.02$, $g_h=0.3\pm0.1$, $\theta=0.4\pm 0.1$ and with an Overhauser field standard deviation of $10.5$ mT$\pm 1.0$ mT (see Fig. 3 of the main text and also Fig. \ref{Fig: Conditional probability HV2} of this document). For all our simulations, we averaged using at least 1000 randomly sampled Overhauser field vectors and observed that all curves converged when further increasing the number of samples.

{We find that the linear polarized conditional probabilities are underestimated when the polarisation trajectory is close to $\ket{H}$. This problem was exacerbated when using a pure dephasing model for the ground state spin dephasing due to the initial sharp exponential decay of coherence, which suggests that a more detailed model of the fluctuating Overhauser field could match the observations better. This small discrepancy may also be explained by the slightly-polarized cavity Purcell effect, or by heavy/light-hole mixing that can alter the polarisation selection rules.}

{From the simulations, we also notice that the optimal $t_{12}$ time depends on the choice of linear excitation polarisation. This is because there is an interplay between the linear excitation polarisation angle and the excited state Zeeman splitting that can alter the effective precession time involved in the time-averaged polarisation measurements. The linearly-polarized excitation will imprint a phase on the initial excited state that dictates how the state evolves due to the Zeeman splitting, and before eventual decay. Thus, the contribution from the hole spin excited state precession to the apparent total average precession time depends on the choice of linear excitation.}


\subsection{Spin-photon correlations and the process map}

An additional benefit of using conditional evolution is that we have direct access to the state of the spin conditioned on the detection of a particular polarized photon. Thus, we can also simulate the process map without integrating spin-photon correlation functions over time. To do so, we first compute the conditional state after a single pulse $\rho_{\vec p}(t)=\mathcal{B}_{\vec p}(t,0)\mathcal{P}_\theta\rho(0)$. Then, we take the expectation value of the electron Pauli operators for the conditional state $\langle\sigma_i^{(e)}(t|\vec p)\rangle=\text{Tr}\{\sigma_i^{(e)}\rho_{\vec p}(t)\}$, where $i\in\{I,x,y,z\}$, $\sigma_I^{(e)}$ is the identity matrix, and $(t|\vec p)$ indicates the expectation value of the state at time $t$ given that a photon of polarisation $\vec p$ was detected between time 0 and time $t$. Then, we have the spin-photon correlations
\begin{equation}
\begin{aligned}
\langle \sigma_i^{(e)}(t)\sigma_I^{(p)}(t)\rangle &= \frac{\langle\sigma_i^{(e)}(t|L)\rangle+\langle\sigma_i^{(e)}(t|R)\rangle}{P_L(t)+P_R(t)}\\
\langle \sigma_i^{(e)}(t)\sigma_x^{(p)}(t)\rangle &= \frac{\langle\sigma_i^{(e)}(t|H)\rangle-\langle\sigma_i^{(e)}(t|V)\rangle}{P_H(t)+P_V(t)}\\
\langle \sigma_i^{(e)}(t)\sigma_y^{(p)}(t)\rangle &= \frac{\langle\sigma_i^{(e)}(t|D)\rangle-\langle\sigma_i^{(e)}(t|A)\rangle}{P_D(t)+P_A(t)}\\
\langle \sigma_i^{(e)}(t)\sigma_z^{(p)}(t)\rangle &= \frac{\langle\sigma_i^{(e)}(t|L)\rangle-\langle\sigma_i^{(e)}(t|R)\rangle}{P_L(t)+P_R(t)},
\end{aligned}
\end{equation}
conditioned on the detection of at least one photon prior to time $t$. Note that the polarisation Pauli operators $\sigma_i^{(p)}(t)$, where $i\in\{I,x,y,z\}$ and $\sigma_I^{(p)}(t)$ is the identity matrix, act on the effective photonic polarisation computational basis $\{\ket{R},\ket{L}\}$ and should not be confused with the continuous set of field operators $a_{\vec p}(t)$ that act on an instantaneous temporal mode of polarisation $\vec p$.

To construct the process map, we simulate the set of 16 correlation functions $\langle \sigma_i^{(e)}(t)\sigma_j^{(p)}(t)\rangle_k$ for four initial spin states $\ket{\psi_k}\in\{\ket{\downarrow},\ket{\uparrow},\ket{+},\ket{+i}\}$. Then, by solving $\text{Tr}\{\sigma^{(e)}_i(t)\sigma^{(p)}_j(t) C_t(\ketbra{\psi_k}{\psi_k})\}=\langle \sigma_i^{(e)}(t)\sigma_j^{(p)}(t)\rangle_k$, we obtain the 64 elements of the process map $C_t$ that describes the spin-photon CNOT gate in addition to the spin precession up until time $t$, provided that a photon was emitted prior to time $t$. By design, $C_t$ applied to an initial quantum dot spin state constructs an effective spin-photon state that reproduces the time-averaged correlation functions that would be measured after a single entanglement cycle. However, the microscopic definition of the effective photonic computational basis in terms of the continuous field modes is not captured by this process map. Thus, although the master-equation model can account for qualities such as the indistinguishability or the temporal profile of the polarisation-entangled photons, the simulated process map only describes how these qualities affect the effective computational basis and it cannot correctly predict how the generated photons will interfere with each other, or any other states of light. To acquire that level of detail, one could use a collisional model to obtain the exact photonic temporal wavefunction emitted by the spin-photon interface \cite{maffei2022energy}, and then average the Overhauser field to obtain the full temporal density matrix of the cluser state.

\begin{figure}
	\includegraphics[width=8cm]{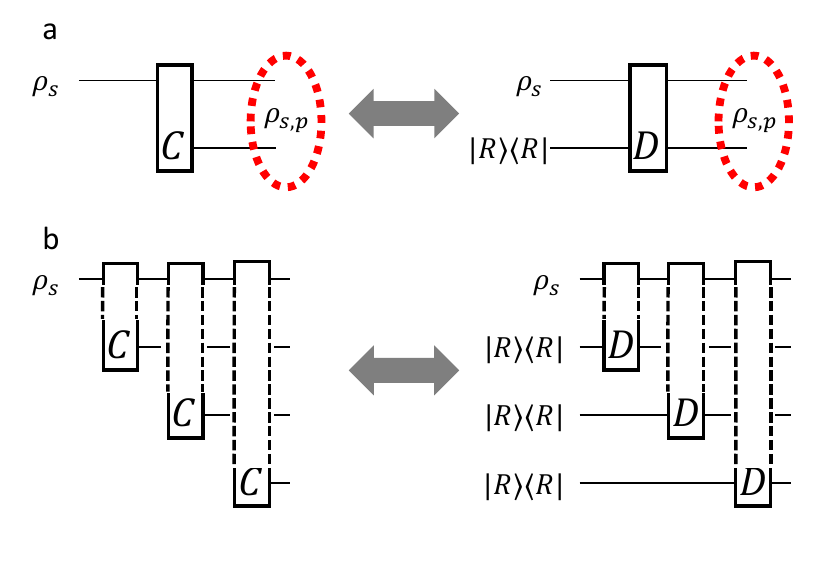}
	\caption{a. Conversion of the LR step process $C(\rho)$ (which creates a photonic qubit) into an equivalent two-qubit process $D(\rho)$. b. Applications for a $k$ repetition of the LR protocol.}
	\label{Fig: process}
\end{figure}

{Note that this process involves the creation of a photonic qubit, thus converting a single spin qubit density matrix $\rho_s$ into a two-qubit spin-photon matrix $C(\rho_s)$. In practice, to facilitate its manipulation, we convert it into a two-qubit process which conserves the number of qubits. To do so, we consider pre-existing photons with initial polarisation R and we construct a two-qubit process $D(\rho_{s,p})$ such that (see also Fig.~\ref{Fig: process}.a.):
\begin{equation}
    D(\rho_s \otimes \ketbra{R}{R}) = C(\rho_s).
\end{equation}
This strategy is similar to converting the emission of a photon into a spin-photon CNOT gate, as is usually presented in the literature~\cite{lindner_proposal_2009, schwartz_deterministic_2016}.}

\section{Extraction of spin-$k$-photon fidelity}

Assuming that we have already a density matrix for the state of the spin and $k-1$ photon  $\rho_{k-1}$, we can use this process to create a  density matrix with $k$ photons $\rho_k = C_k(\rho_{k-1})$.
Here, the $k$ photon process $C_k$ is based on the emission and spin rotation process $C(\rho)$ but is also trivially acting on the larger subspace which includes the other photons:
\begin{equation}
    C_k = C\otimes_{i=1}^{k-1} I
\end{equation}
where we apply $C$ on the spin state and the identity $I$ on the photons.

Therefore, a spin+$k$ photon density matrix is produced by applying $k$ times $C(\rho)$ on the initial spin $\rho_s$ which translate into:
\begin{equation}
\rho_{n} = C_k \circ C_{k-1} \circ ... \circ C_{1} (\rho_s)
\end{equation}

Similarly, we can also use the equivalent process $D(\rho)$ as shown in Fig.~\ref{Fig: process}.b.
We can compare this states to the ideal state spin+$k$ photon state given by applying the ideal map $\tilde C_k$ defined the same way as $C_k$, but using the ideal emission + spin precession process $\tilde C(\rho)$.
The spin+$k$ photon fidelity is thus calculated by comparing the resulting density matrix $\rho_k$, with the ideal state $\tilde \rho_k$:
\begin{equation}
F_k = \Tr[\tilde \rho_k \rho_k].    
\end{equation}



\end{document}